\documentclass[twocolumn,showpacs,preprintnumbers,amsmath,amssymb]{revtex4}
\usepackage{graphicx}
\usepackage{dcolumn}
\usepackage{bm}

\begin{document}

\title{Cohesive energies of Fe-based glass-forming alloys}

\author{M. Mihalkovi\v{c}}
 \altaffiliation[Also at ]{Institute of Physics, Slovak Academy of Sciences,
84228 Bratislava, Slovakia}
\author{M. Widom}
\affiliation{
Department of Physics\\
Carnegie Mellon University\\
Pittsburgh, PA  15213
}

\date{\today}

\begin{abstract}
We calculate the cohesive energies of Fe-based glass-forming alloys in
the B-Fe-Y-Zr quaternary system.  Our {\it ab-initio} calculations
fully relax atomic positions and lattice parameters yielding
enthalpies of mixing at T=0K.  We examine both the known equilibrium
and metastable phases as well as a selection of plausible structures
drawn from related alloy systems.  This method, generally reproduces
experimentally determined phase diagrams while providing additional
information about energetics of metastable and unstable structures.
In particular we can identify crystalline structures whose formation
competes with the metallic glass.  In some cases we identify
previously unknown structures or observe possible errors in the
experimental phase diagrams.
\end{abstract}

\maketitle

\section{\label{sec:Intro}Introduction}

Calculation of alloy phase diagrams from first-principles is necessary
to achieve the goal of materials by design~\cite{Olson}.  Bulk
metallic glass-forming alloys, which often contain three or more
chemical elements provide a useful test case.  Recently discovered
many-component alloys\cite{BMG_ZrAlNi,BMG_ZrX,BMG_PdNiP} solidify in
amorphous structures at relatively low cooling rates. These materials
display intriguing and potentially useful mechanical properties
including nearly perfect elasticity\cite{Elasticity}. Amorphous
Fe-based alloys are interesting for both their structural and their
magnetic properties.  Achieving bulk glass formation could extend the
range of potential applications of these materials.

To understand factors limiting bulk glass formation, we perform {\it
ab-initio} total energy calculations on the quaternary compound
B-Fe-Y-Zr as well as its binary and ternary subsystems such as B-Fe
and B-Fe-Zr.  Our calculations use the plane-wave electronic density
functional theory program VASP~\cite{VASP,VASP2}. We identify the
crystalline structures whose formation competes with the amorphous
structure of the supercooled liquid. To this end, we calculate the
cohesive energies of stable, metastable and hypothetical crystal
structures throughout the alloy composition space.  Standard
metallurgical databases~\cite{Pearson,BinaryPD,TernaryPD}list known
stable and metastable structures.  Chemically similar alloy systems
provide hypothetical structures to test.

Cohesive energies of stable and metastable phases yield thermodynamic
driving forces for crystallization. The structural complexity of these
phases gives some insight into the possibility of their nucleation and
growth during a rapid quench. Our main results are the identification
of the structure types CFe$_3$ (Pearson symbol oP16) and
C$_6$Cr$_{23}$ (Pearson symbol cF116) as the two main competitors to
the B-Fe glass. Alloying with Zr does surprisingly little to
destabilize the C$_6$Cr$_{23}$ structure, while alloying instead with
Y does reduce stability of this structure. On the other hand, alloying
with Y stabilizes certain other ternary structures. On this basis we
deduce advantageous composition ranges.

Energetically favorable structural motifs identified within these
phases can be compared with structural models of the metallic glass.
We identify a class of Boron coordination polyhedra related to the
trigonal prism~\cite{Gaskell} with some distortions. These polyhedra
may be arranged in many ways, some leading to simple crystal structures
with very low energy but many more whose energies and local structure
closely resemble the glass. We call the entire class of these structures
``amorphous approximants'', by analogy with the concept of approximants to
quasicrystal structures~\cite{QCApprox}.

The following two sections of this paper (sections~\ref{sec:Methods}
and \ref{sec:Results}) present our calculational methods and the
resulting cohesive energy data.  Although we study here Fe-based
glass-forming alloy systems, our basic method can be applied to any
alloy system.  Besides checking known experimental phase diagrams, our
calculations provide energetic information which is often not known
experimentally, especially in the case of metastable and amorphous
structures.  In addition, we propose likely structures for compounds
whose existence was known but whose structures were unknown, for
example, B$_4$Fe$_4$Y, B$_4$FeY and B$_6$Fe$_2$Y$_5$. Conversely, in
some cases our results call into question details of the established
phase diagrams. For example: the claimed high temperature stability of
BZr is most likely only metastability in reality; the Co$_7$Y$_2$
structure may be stable in Fe-Y though it has not been reported; the
phase Fe$_{17}$Y$_2$, related to important permanent magnet
materials~\cite{RFe}, is possibly only a high temperature
phase. Further, we can predict phase diagrams of alloy systems such as
B-Y-Zr, Fe-Y-Zr and B-Fe-Y-Zr that have not been established
experimentally.

Finally, section~\ref{sec:Discuss} analyzes the crystal structures and
correlates their atomic arrangements with their cohesive energies.
Amorphous approximants are presented in subsection~\ref{sec:Approx}.

\section{\label{sec:Methods}Methods}

Our interest in the binary B-Fe and ternary B-Fe-Zr and B-Fe-Y
compounds led to the study of all elemental, binary, ternary and
quaternary combinations of the elements B-Fe-Y-Zr.  We selected
structures for study that are known as stable or metastable structures
in the phase diagrams of these alloy systems or chemically similar
alloy systems. For example we consider known C-Fe structures
(e.g. CFe$_3$.oP16) with B replacing C, etc.  Our notation for
structure type is to first give the {\em prototype} (some familiar
isostructural compound, e.g. CFe$_3$) followed by the {\em Pearson
symbol} (indicating the point symmetry, centering information, and
number of atomic sites per unit cell, e.g. oP16 for
\underline{o}rthorhombic \underline{P}rimitive 16-atom cell).

Our sources for established phase diagrams and structures include
standard references~\cite{Pearson,BinaryPD,TernaryPD}, individual
publications and private communication.  Information from these
sources has been compiled into a database containing over 1000
structures that we search to match criteria such as chemical elements,
stoichiometry, and atomic size ratios.  Some additional structures
examined are liquid and amorphous structures, obtained from {\em
ab-initio} molecular dynamics simulation.

Our {\em ab-initio} calculations use the program VASP (version 4.5.5)
together with the projector-augmented wave method, an all-electron
generalization of the pseudopotential approach~\cite{PAW,KJ_PAW}.  We
employ the Perdew-Wang generalized gradient approximation~\cite{PW91}
(GGA) exchange-correlation functional with the Vosko-Wilkes
Nussair~\cite{VWN} spin interpolation. These choices give excellent
results for bulk elemental Fe~\cite{KJ_PAW}. GGA is needed instead of
LDA is necessary to properly reproduce magnetization and lattice
constants~\cite{Moroni}. Our magnetic calculations are spin-polarized
(i.e. collinear magnetization) and are employed for any structure
containing 50\% Fe or higher.

VASP solves for the self-consistent electronic structure in reciprocal
space, using a plane wave basis. It requires that we choose the
reciprocal space grid appropriately and demonstrate convergence in the
number of k-points used and the plane wave energy cutoff. We construct
k-point grids whose spacing is nearly isotropic in reciprocal
space. Mostly we use Monkhorst-Pack grids, although for hexagonal
structures we use Gamma-centered grids. Our k-point density is
sufficient that all structural energies are converged to a precision
of 10 meV/atom or better. All energies for structures that lie on or
near the convex hull are converged to a precision of 1 meV/atom or better.

In general our relaxations allow variations of cell volume and shape,
as well as atomic displacements, consistent with the symmetry of the
starting structure.  Relaxations run until an accuracy of 1meV/atom or
better is reached.  During relaxation we use Methfessel-Paxton
Fermi-surface smearing with width 0.2 eV (the VASP default
choice~\cite{VASP_home}.  When smearing is employed we report the {\em
energy} (extrapolated to zero smearing), not the fictitious {\em free
energy}. For many structures on or near the convex hull, we
recalculated the energy of our best-relaxed structure using the
tetrahedron method without smearing.  This test confirms we reached
our 1 meV/atom precision goal.

Table~\ref{tab:converge} demonstrates convergence with respect to
k-point grid and plane wave cutoff energy, reporting the energies of
two metastable variants of Fe$_3$B, one with structure type
Fe$_3$C.oP16 and one with structure type Ni$_3$P.tI32.  We see how
energies $E$ and energy differences $\Delta E$ converge as the k-point
mesh grows at fixed energy cutoff.  The energy difference $\Delta E$
at low precision has even the incorrect sign, while medium and high
precision agree to within the desired 1meV/atom accuracy. For the
results presented below we employ a constant energy cutoff of 319 eV,
consistent with medium precision.

Certain calculated structural quantities can be compared directly with
experiment. For example, for Fe$_3$B.oP16 we obtain converged volume
9.78~\AA$^3$/atom, $b/a=1.23$ and $c/a=1.52$ compared with the
experimental values, respectively, of 10.09~\AA$^3$/atom, 1.22 and
1.51. Likewise, for Fe$_3$B.tI32 we find volume 9.61~\AA$^3$/atom and
$c/a=2.03$ compared with experimental values 10.06~\AA$^3$/atom and
2.01.  Our underestimation of the volume reflects both thermal
expansion (experimental volumes are at room temperature, while our
calculations are for T=0K ground states) and known systematic errors
associated with density functional theory.

\begin{table}
\caption{\label{tab:converge}Convergence of magnetization and energy
with respect to k-point density and plane wave cutoff. Numbers in
left-hand column are the product of lattice parameter times the number
of k-point subdivisions. Values of $E_c$ are plane wave cutoff energies
in eV}
\begin{tabular}{|c|c||r|r||r|r||r|}
\hline
\multicolumn{2}{|c||}{Param} & 
\multicolumn{2}{c||}{Fe$3$C.oP16} &
\multicolumn{2}{c||}{Ni$_3$P.tI32}& \\
\hline
grid & $E_c$ &  M   &    E    & M    & E       & $\Delta  E$ \\
\hline
1    & 239   & 1.38 & -8.1227 & 0.37 & -8.1073 &  0.0155 \\
10   & 239   & 1.16 & -7.9698 & 0.80 & -7.9283 &  0.0415 \\
20   & 239   & 1.29 & -8.0460 & 1.33 & -8.0519 & -0.0059 \\
28   & 239   & 1.22 & -7.9870 & 1.27 & -7.9899 & -0.0029 \\
36   & 239   & 1.20 & -7.9629 & 1.26 & -7.9656 & -0.0027 \\
44   & 239   & 1.20 & -7.9554 & 1.26 & -7.9572 & -0.0017 \\
52   & 239   & 1.20 & -7.9525 & 1.25 & -7.9543 & -0.0018 \\
\hline
1    & 319   & 1.38 & -8.1725 & 0.82 & -8.0873 &  0.0852 \\
10   & 319   & 1.51 & -8.0683 & 1.33 & -8.0365 &  0.0318 \\
20   & 319   & 1.49 & -8.0430 & 1.38 & -8.0428 &  0.0002 \\
28   & 319   & 1.48 & -8.0441 & 1.39 & -8.0420 &  0.0020 \\
36   & 319   & 1.49 & -8.0437 & 1.39 & -8.0409 &  0.0028 \\
44   & 319   & 1.50 & -8.0437 & 1.39 & -8.0410 &  0.0027 \\
52   & 319   & 1.50 & -8.0437 & 1.39 & -8.0411 &  0.0026 \\
\hline
1    & 398   & 1.38 & -9.1688 & 1.43 & -8.0991 &  0.0696 \\
10   & 398   & 1.50 & -8.0667 & 1.51 & -8.0667 &  0.0334 \\
20   & 398   & 1.48 & -8.0382 & 1.38 & -8.0383 & -0.0001 \\
28   & 398   & 1.48 & -8.0390 & 1.39 & -8.0372 &  0.0018 \\
36   & 398   & 1.48 & -8.0386 & 1.38 & -8.0362 &  0.0025 \\
44   & 398   & 1.49 & -8.0386 & 1.39 & -8.0362 &  0.0024 \\
52   & 398   & 1.49 & -8.0386 & 1.39 & -8.0363 &  0.0023 \\
\hline
\end{tabular}
\end{table}

The composition space of an $N$-component alloy is a set of $N$
composition variables $\{x_i: i=1, 2,..., N\}$ obeying $\sum_{i=1}^N
x_i=1$. The set forms an $N-1$ dimensional simplex (respectively, a
point, line segment, triangle and tetrahedron for $N=1, 2, 3, 4$).
Structural energies form a scatter-plot over this simplex.  Stable low
temperature phases lie on vertices of the convex hull of the energy
versus composition scatter-plot. Edges and facets of the convex hull
represent coexistence regions of the phases at adjoining vertices.
Lines and triangles joining low temperature phases phases will be
referred to as tie-lines and tie-planes. A tie-surface in general
refers to the hyperplane joining $N$ or fewer points in the
$N$-component energy scatter-plot.

The tie-surface connecting all pure elements in their lowest energy
structures forms a useful reference for alloy energies. The distance
$\Delta H_{for}$ of an alloy energy from the tie-surface joining pure
elements is known as its enthalpy of formation (enthalpy because
volume relaxation means we work at fixed pressure, $P=0$). Strong
compound formation is reflected in large negative enthalpy of
formation.

High temperature phases should lie above the convex hull, but be
sufficiently close that entropic effects (e.g. phonons or chemical
substitution) can stabilize them. Metastable phases also should lie
close to the convex hull, so that their free energy is less than the
liquid free energy at temperatures below freezing. Although $\Delta
H_{for}$ is usually negative for high temperature and metastable
phases, their energy difference $\Delta E$ from the convex hull is
small and positive. The value of $\Delta E$ is a measure of the
thermodynamic driving force for decomposition into the appropriate
combination of stable phases.

Using these methods, we built a database of structural energies.  For
a given $N$-component alloy system of interest we extract from our
database energies of structures containing all, of just some, of the
chosen elements. We use a standard convex hull program
(qhull~\cite{qhull}) to identify stable structures and the coexistence
regions that connect them. Based on the output of this program, we
calculate values of $\Delta H_{for}$ and $\Delta E$ for every
structure.

Our methods introduce systematic errors associated with the PAW
implementation and even with the underlying density functional
theory. Provided these errors vary smoothly with composition, the
identity of convex hull vertices will not be affected in most
cases. However, the tie-lines, tie-planes, etc. grow progressively
more sensitive to error. It is probable that even when we correctly
identify the stable phases, we may misidentify their coexistence
regions.

Previous workers carried out analogous studies, though mainly on
binary alloys.  Hafner~\cite{HafnerHam} wrote a general introduction
to the subject of {\em ab-initio} alloy phase diagram
prediction. Miedema, de Boer and coworkers~\cite{deBoer} performed
extensive semi-empirical analysis of binary alloy systems, including
almost all binary alloys of Fe~\cite{Boom}.
Others~\cite{Foiles,Sluiter,Wolverton} take more rigorous approaches,
sometimes including finite temperature effects of vibrational and
configurational entropy. Many cohesive energy calculations have been
collected in online databases~\cite{HLS,NIMS,MS}.  All the data we
present here, and a great deal more, can be found on the
WWW~\cite{alloy_home}.

\section{\label{sec:Results}Results}

The $N=4$-component B-Fe-Y-Zr quaternary alloy system contains many
subsystems: four pure elements, six binary alloy systems and four
ternaries. This section presents our results in order of increasing
number of components.  We adopt alphabetical order in naming alloy
systems, because this brings some order to the proliferation of
chemical combinations in multicomponent systems.

In most cases we reproduce the known equilibrium and metastable phase
diagrams with surprising accuracy.  In a few cases lingering
discrepancies may reveal limitations of our method or cast doubt on
the accepted phase diagrams.  In some cases where the existence of a
phase was known experimentally but not its structure we suggest
probable structures.  Our quaternary and two of our ternary phase
diagrams have never been determined experimentally.

\subsection{\label{sec:elements}Pure elements}

Each of the four elements under study exhibits solid-solid phase
transitions in addition to its melting transition, and verifying the
relative energies of the different structures is a nontrivial test of
our calculational method.

The precise low temperature structure of Boron is unknown. The presumed
equilibrium phase, designated $\beta$ is rhombohedral with
approximately 108 atoms per unit cell.  Atomic positions are
known, but there is partial occupancy and probably strong correlations
among the partially occupied sites, that have not been adequately
resolved.

The basic structure of $\beta$-Boron (denoted B.hR105) consists of
overlapping B$_{156}$ clusters with icosahedral symmetry, located at
the vertices of the primitive rhombohedral cell~\cite{Hoard}.  The
$12*(12+1)=156$-atom cluster assembles twelve 12-atom icosahedra
surrounding one central icosahedron.  All icosahedra all empty.  A
single extra atom at the body center of the rhombohedral cell is not
part of any B$_{156}$ cluster.  An alternate description concentrates
on non-overlapping B$_{84}$ clusters, obtained by removing the outer
halves of the 12 outer icosahedra of B$_{156}$.  In addition to the
B$_{84}$ cluster the basic structure contains two B$_{10}$ clusters
connected by the extra B atom in the center of the rhombohedral
primitive cell, yielding 105 atoms per unit cell.

Wyckoff positions of B.hR105 are labeled B1-B15, with the B15 site the
cell center.  Structural refinements~\cite{Hoard,Hughes,Callmer,Slack}
find the B13 sites surrounding the B15 atom at the center of the
rhombohedral cell only fractionally occupied, while an extra B16 Boron
atom approximately compensates the missing electron density.  The
refinement by Slack~\cite{Slack} additionally reports five other Boron
sites with small occupancy factors, and proposes tentative a model for
occupancy correlations.  The Pearson symbol for the Slack model is
hR141.

Our calculations show the basic B.hR105 structure higher in energy
than the presumed metastable $\alpha$-B structure (B.hR12) by 25
meV/atom.  We find, in agreement with Slack~\cite{Slack} but contrary
to naive assumption, that the choice of which B16 atom to insert is
not correlated to the B13 site occupancy. The B16 atom lowers the
total energy by as much as 8 eV per primitive rhombohedral cell,
narrowing the energy difference between $\alpha$ and $\beta$ boron to
$\sim$11 meV/atom.  Insertion of a second B16 atom, or other more
complex modifications, are second-order corrections to the total
energy.  Our best model of the fractional occupancies additionally
replaces one B13 atom by a new B19 atom (106 atoms per primitive
cell), provided the B19 atom is not a nearest neighbor of the B16
atom.  This model remains unstable relative to $\alpha$-B by $\sim 5$
meV/atom.

Pure elemental Iron passes through four solid phases,
$\alpha$-$\delta$, as temperature rises from T=0K to melting. $\alpha$
is BCC (Pearson cI2) and ferromagnetic. At its Curie temperature it
transforms to $\beta$ which is also BCC. At higher temperatures it
transforms to $\gamma$, which is FCC (Pearson cF4) and finally to
$\delta$, again BCC, before melting.  First-principles
calculations~\cite{gamma-Fe} show that the magnetic ground state of
$\gamma$-Fe is a noncollinear antiferromagnet, while at the
experimental atomic volume it is a collinear
antiferromagnet~\cite{AFMD}.  According to our calculations, the
relaxed energy of this collinear antiferromagnetic structure is 81
meV/atom above the energy of $\alpha$-Fe.

The low temperature $\alpha$ phases of Yttrium and Zirconium are both
hexagonal (hP2) and their high temperature $\beta$ phases are both BCC
(cI2).  Our calculations agree with these facts.

\subsection{\label{sec:binaries}Binaries}
\subsubsection{\label{sec:Fe-B}Fe-B}

The established Fe-B phase diagram contains just two compounds,
Fe$_2$B.tI12 and BFe.oP8, each with simple structures. A number of
metastable phases exist in the Fe-rich end, namely CFe$_3$.oP16,
Ni$_3$P.tI32 and C$_6$Cr$_{23}$.cF116. These occur close to the deep
eutectic at 17~\%~B, and hence are important competitors to glass
formation.

\begin{figure*}
\includegraphics*[width=7in]{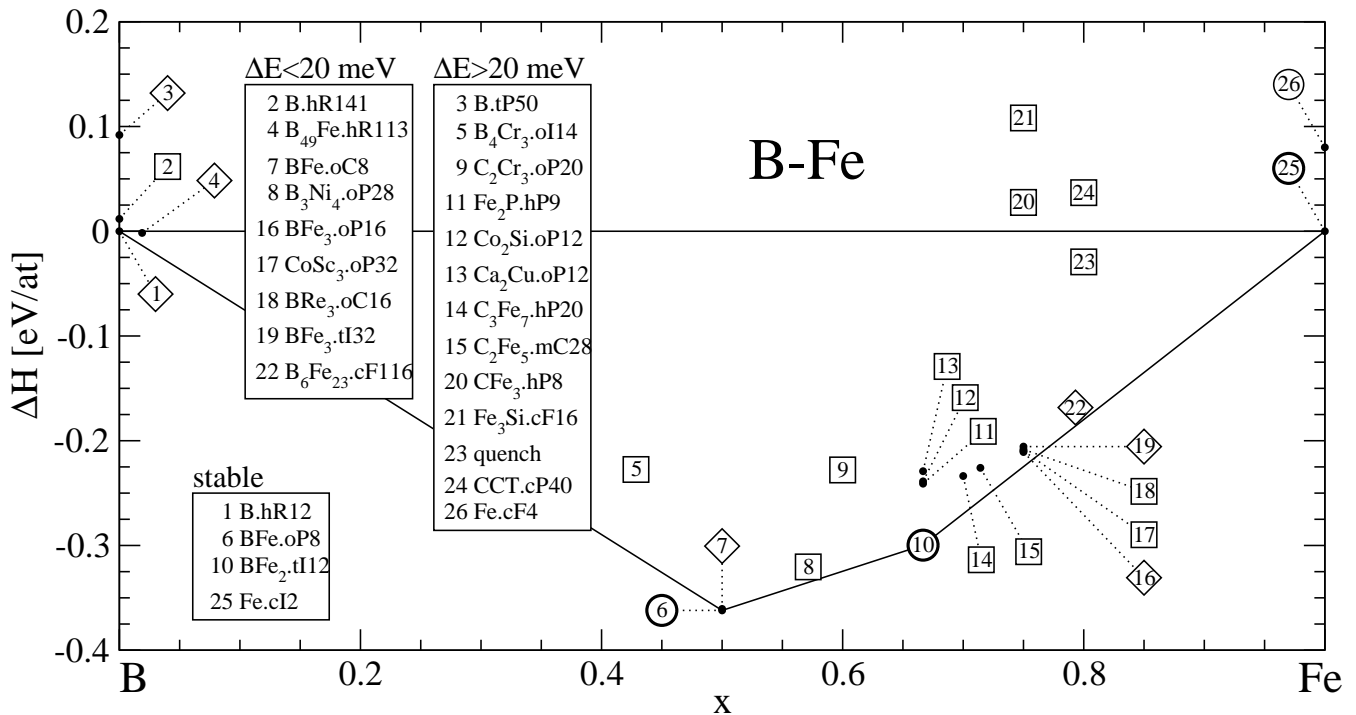}
\caption{\label{fig:b-fe} Enthalpies of formation and their convex
hulls for the B-Fe binary alloys. Notation: heavy
circles denote known low temperature phases, light circles denote
known high temperature phases, diamonds denote known metastable
phases, squares denote unreported structures.}
\end{figure*}

Our calculations shown in Fig.~\ref{fig:b-fe} reproduce the
established phase diagram perfectly, with the known stable phases
lying on the convex hull and the known metastable phases lying within
20 meV/atom above (at the eutectic melting temperature of T=1447 K, a
characteristic thermal energy is $k_B T=125$ meV/atom).  Structures
occurring as stable or metastable phases in other alloy systems (but
not in Fe-B) lie further above the convex hull, except for the
BRe$_3$.oC16 and CoSc$_3$.oP32 structures. On the basis of our
calculation, BRe$_3$ and CoSc$_3$ {\em should} occur as metastable
phases in Fe-B, although they have not been reported.

A prior {\em ab-initio} study of the Fe-B structures investigating
magnetism and bonding~\cite{Ching} reported a covalent character to
Fe-B and B-B bonds while the Fe-Fe bonds are metallic in nature,
together with a significant charge transfer from Fe to B. Spin
polarization of the B atoms is weak and {\em opposite} to the Fe
atoms.  Our calculations support those conclusions.

Given our perfect agreement with the established phase diagram, it is
surprising that our calculated enthalpies (respectively -368 and -308
meV/atom for FeB and Fe$_2$B) differ greatly from measured values
(respectively -676 and -707 meV/atom for FeB and Fe$_2$B~\cite{LB}).
This may be due in part to the fact that our calculation are
performed at T=0K while the measurements were done at T=1385K.  Pure
elemental Fe undergoes two phase transitions (one structural, one
magnetic) as temperature drops, which could be partly
responsible for this discrepancy.

One hypothesis on the glass-forming ability of Fe-B is that very
simple, easy to nucleate and grow crystal structures, are destabilized
by the size contrast~\cite{Egami} of Fe (nominal diameter=2.48\AA) and
B (nominal diameter=1.80\AA).  We observe this principle in action in
the Fe-rich end of this phase diagram.  Consider a substitutional solid
solution of Fe and B, around composition BFe$_3$, based on the BCC
structure of Fe.  The Fe$_3$Si.cF16 structure is a particular
realization of such a structure, in which the Fe and B atoms arrange
at regular positions.  However, the energy of cF16 is much higher than
the metastable oP16 structure, which can be reached through distortion
of the cF16 lattice.  It seems that the lattice strain caused by size
mismatch destabilizes the BCC solid solution, converting cF16 into
oP16.

We note in addition that B will not stabilize the FCC structure of Fe
by substitution to form the AuCu3.cP4 structure, though such stabilization
does occur with larger atoms, for example Fe$_3$Ge.

Alternatively, B might enter as an interstitial, as indeed C enters
into FCC lattices of Fe in octahedral or tetrahedral sites
(respectively, in the metastable Fe$_3$C.hP8 and Fe$_4$C.cP5
structures).  However, owing to the slightly larger size of B compared
with C (nominal diameter=1.43\AA), these structures are far above the
convex hull in the B-Fe energy scatter plot.

It seems that only fairly complicated crystal structures exist near
compositions of about 25\% B.  The difficulty of nucleating and
growing these structures may aid in glass formability.  An estimate of
the thermodynamic driving force for nucleation can be obtained by
comparison of three energy scales.  The metastable structures are about
10-20 meV/atom above the tie-line joining Fe$_2$B to $\alpha$-Fe.
Further details of Fe-B structures, especially focusing on B-atom
environments and the occurrence of trigonal prism structures is given
in the discussion section~\ref{sec:Discuss}.

A quenched amorphous structure at composition Fe$_{80}$B$_{20}$ is
about 170 meV/atom above the tie-line, and the liquid is 350 meV/atom
above.  Candidate liquid and amorphous structures were produced by
liquid state molecular dynamics on 100 atom samples using Nos\'e
dynamics at temperature T=1500K, then quenched by conjugate gradient
relaxation.  Atomic displacements under relaxation are large, averaging
about 0.6~\AA. The liquid runs were at fixed volume, 7\% greater than
the volume of the quenched amorphous samples. All molecular dynamics
and quenching runs were done with spin polarization, using the
$\Gamma$ k-point only.

\subsubsection{\label{sec:Fe-Zr+Fe-Y}Fe-Zr and Fe-Y binaries}

A promising glass-forming strategy is to start with a good
glass-forming binary such as B-Fe, then add one or more additional
elements to further destabilize any crystalline structures. Large
atoms such as Zr (nominal diameter=3.18\AA) and Y (nominal
diameter=3.55\AA) are promising because they associate well with Fe
and B but they differ strongly in size from either Fe or B. Before
turning to these ternary and quaternary compounds, we briefly examine
the Fe-Zr and Fe-Y binaries.

\begin{figure*}
\includegraphics*[width=7in]{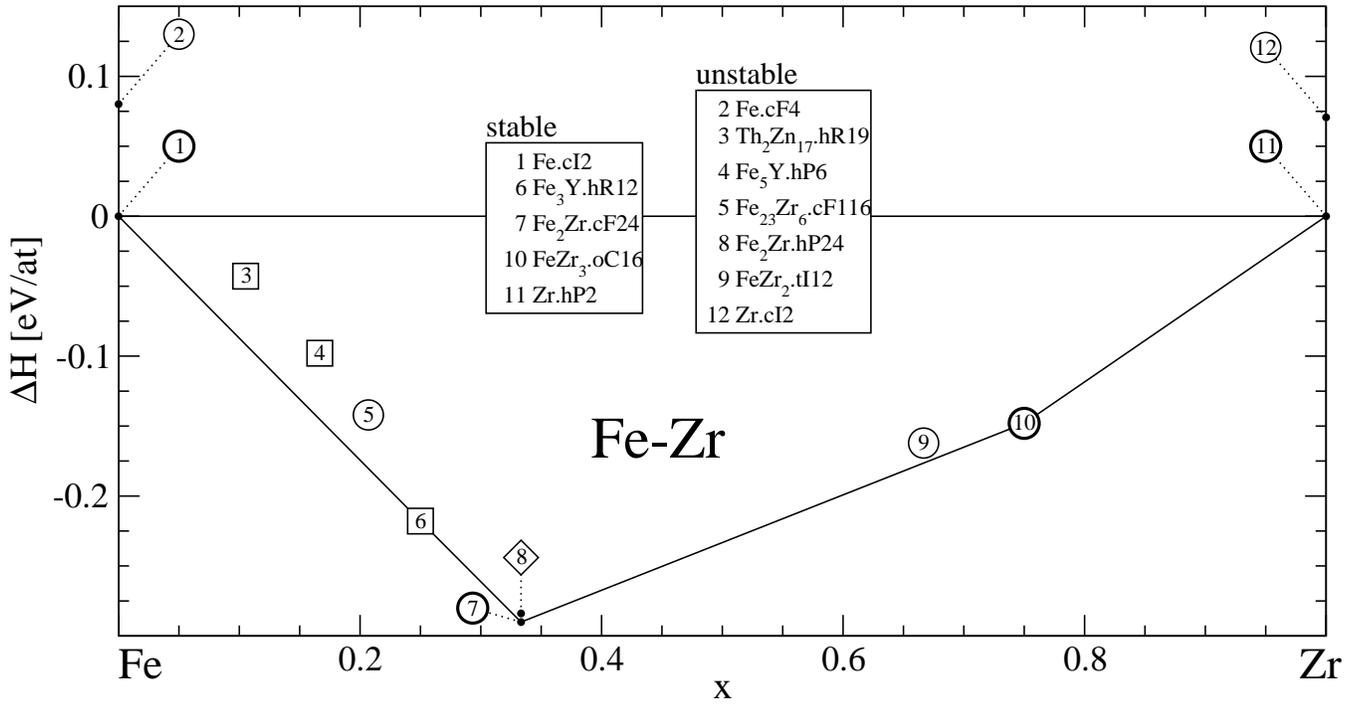}
\caption{\label{fig:fe-zr} Fe-Zr alloy enthalpies. Plotting symbols as in
Fig.~\ref{fig:b-fe}}
\end{figure*}
\begin{figure*}
\includegraphics*[width=7in]{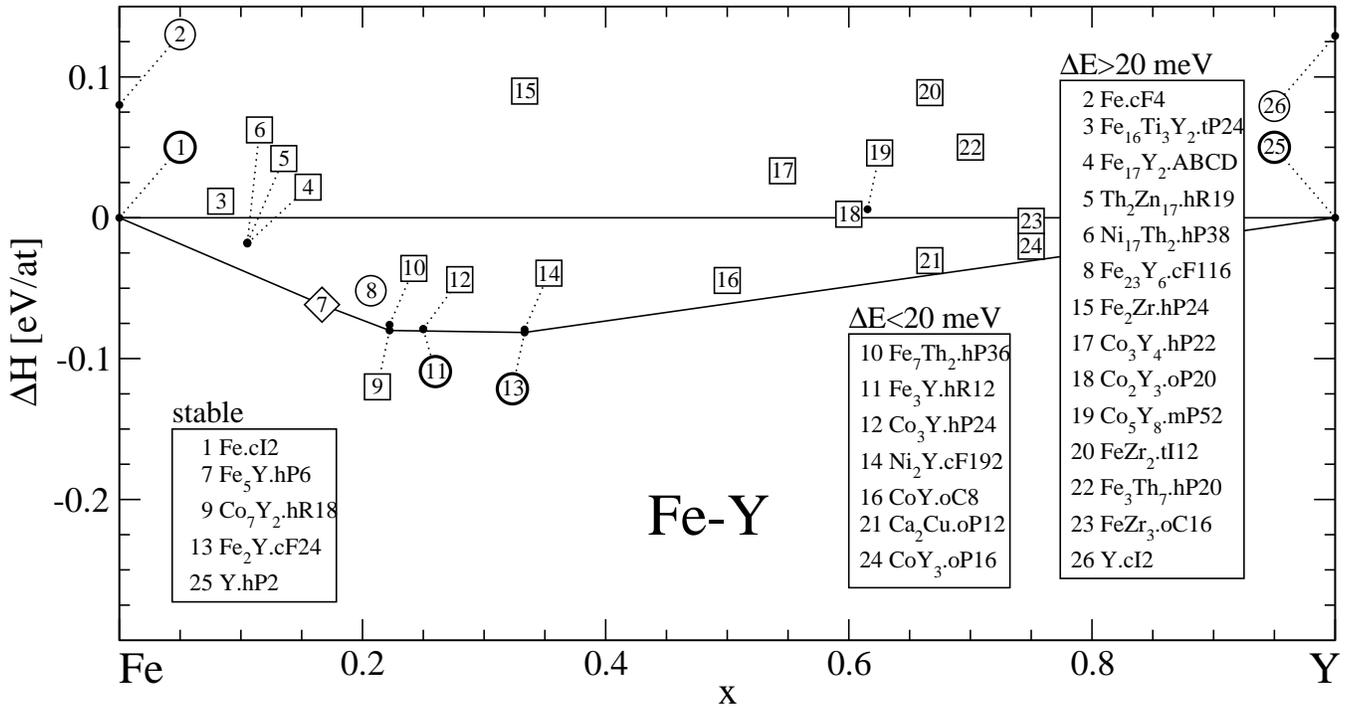}
\caption{\label{fig:fe-y} Fe-Y enthalpies.  Plotting symbols as in
Fig.~\ref{fig:b-fe}}
\end{figure*}

The established Fe-Zr phase diagram~\cite{PD_new} contains three low
temperature compounds, Fe$_2$Zr.cF24, FeZr$_2$.tI12 and FeZr$_3$.oC16,
one high temperature compound Fe$_{23}$Zr$_6$.cF116 (this cF116
structure is quite distinct from the C$_6$Cr$_{23}$.cF116 structure
which confusingly shares the same stoichiometry and Pearson symbol)
and one metastable compound Fe$_2$Zr.hP24.  Our calculation (see
Fig.~\ref{fig:fe-zr}) is in excellent agreement with experiment. Every
known low temperature phase lies on the convex hull, and the high
temperature and metastable compounds lie close above it.

One unknown structure, Fe$_3$Y.hR12, appears on the convex hull where
no stable compound is known experimentally. Most likely this reflects
an inaccuracy of our methods. Because its stability relative to the
tie-line joining Fe$_2$Zr to pure Fe is about 1 meV/atom, small errors
(either calculational or arising from approximations of density
functional theory) could account for this difference.  Alternatively,
the phase could truly be stable, but hard to observe experimentally
because the driving force for its formation is weak. This matter
requires further theoretical and experimental analysis, but for the
study of glass formation all we care about is that its energy lies
close to the tie-line from Fe$_2$Zr to $\alpha$-Fe.

\begin{figure*}
\includegraphics*[width=7in]{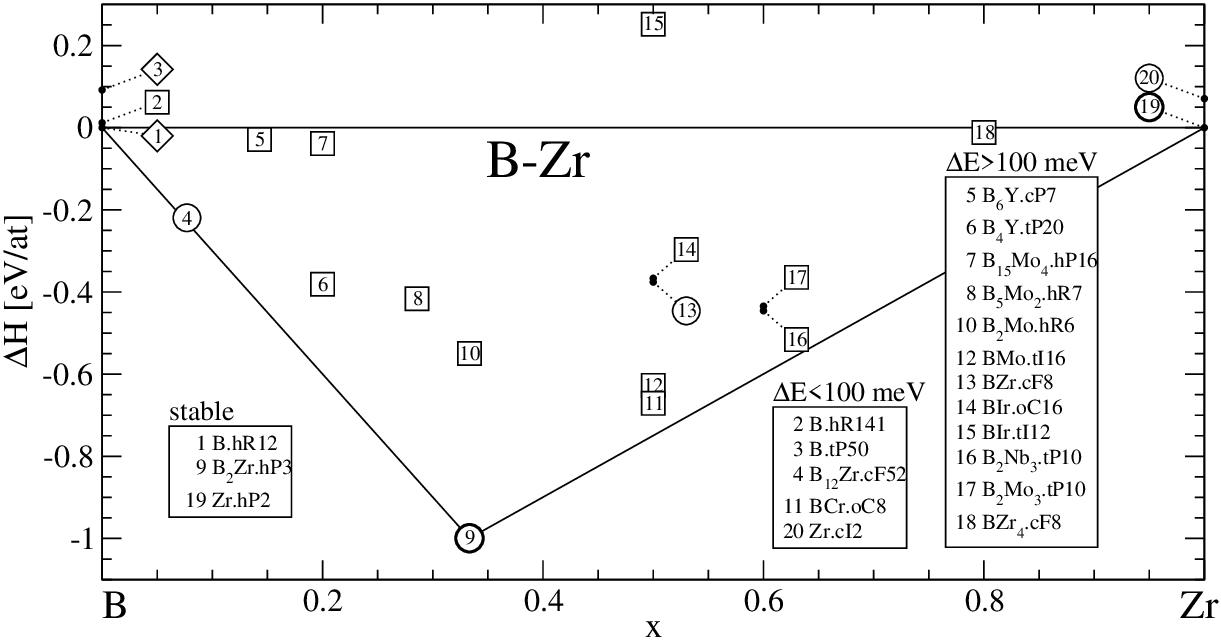}
\caption{\label{fig:b-zr} B-Zr enthalpies.  Plotting symbols as in
Fig.~\ref{fig:b-fe}}
\end{figure*}
\begin{figure*}
\includegraphics*[width=7in]{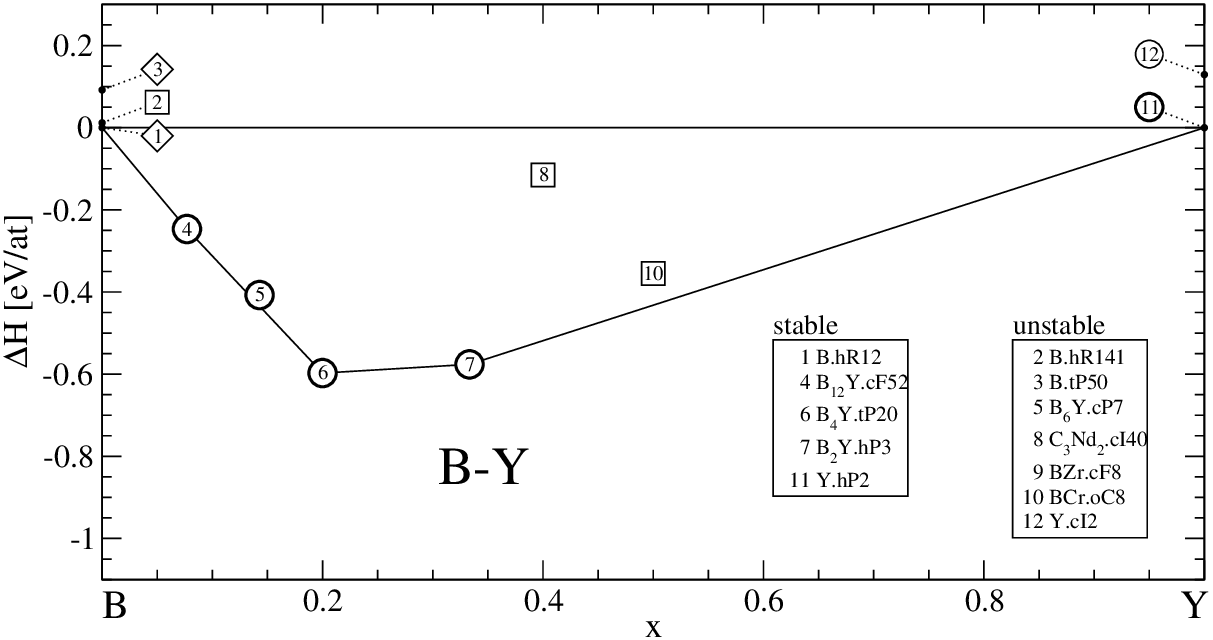}
\caption{\label{fig:b-y} B-Y enthalpies.  Plotting symbols as in
Fig.~\ref{fig:b-fe}}
\end{figure*}

Now consider Fe-Y. The established phase diagram~\cite{PD_new}
contains all the same phases as Fe-Zr and two additional phases,
Fe$_3$Y.hR12 and Fe$_{17}$Y$_2$.  Certain features of the phase
diagram are thermodynamically improbable~\cite{improb,improb2}.  The
close proximity of line compound Fe$_{23}$Y$_6$.cF116 (at low
temperature) to the line compound Fe$_3$Y.hR12 is highly unlikely. We
presume that Fe$_{23}$Y$_6$.cF116 is stable at high temperatures only,
where the phase diagram shows a broad composition range. The strong
asymmetry of the liquidus of Fe$_3$Y.hR12 also is improbable but we
have no proposed alternative at present.

\begin{figure*}
\includegraphics*[width=7in]{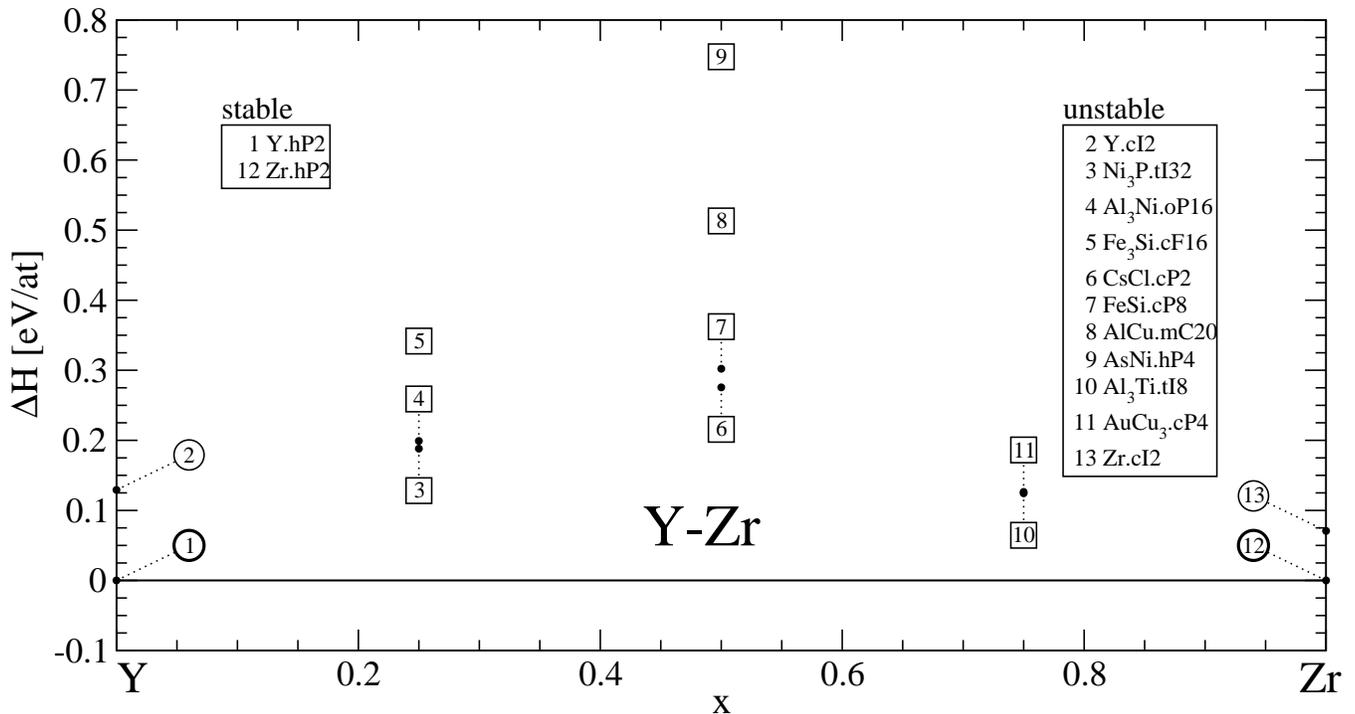}
\caption{\label{fig:y-zr} Y-Zr enthalpies. Plotting symbols as in
Fig.~\ref{fig:b-fe}.}
\end{figure*}

The structure of Fe$_{17}$Y$_2$ has not been properly identified, and
is believed to occur in at least two variants. We follow the lead
of Massalski and Okhamoto and identify these variants as
Th$_2$Zn$_{17}$.hR19 (low temperature) and Ni$_{17}$Th$_2$.hP38 (high
temperature).  Other reported variants~\cite{Pearson} of this phase
are Fe$_{17}$Ho$_2$.hP44 and Ni$_{19}$Th$_2$.hP80.  The occurrence of
several structural variants, most with partial occupancy, suggest a
possible entropic stabilization mechanism by structural disorder.
See the synopsis of the basic structure and its degrees of freedom
in the discussion section~\ref{sec:Discuss}.

Our calculations for Fe-Y (Fig.~\ref{fig:fe-y}) present certain
disagreements in comparison with experiment. Notably, all variants of
Fe$_{17}$Y$_2$ lie above the convex hull and thus are predicted as
high temperature or metastable. Meanwhile, Fe$_5$Y.hP6, believed
metastable, touches the convex hull as does the unreported structure
Co$_7$Y$_2$.hR18.

For the failure of Fe$_{17}$Y$_2$ to meet the convex hull, three
possible explanations are: (1) The structure of Fe$_{17}$Y$_2$ has not
yet been correctly determined (we mentioned this above and discuss it
further in section~\ref{sec:Fe17Y2}); (2) Our calculations are
seriously flawed and unable to properly compare the energies of Fe-Y
compounds (we checked that changes in cutoff energy, pseudopotentials
and exchange-correlation functional have no significant impact); (3)
Fe$_{17}$Y$_2$ is only metastable or high temperature and the true low
temperature state is a coexistence of Fe$_5$Y and pure Fe.

A simple mechanism to explain high temperature stability is note that
the energy of Fe$_{17}$Y$_2$ lies {\em below} the tie-line from
Fe$_5$Y to $\gamma$-Fe.  Because $\gamma$-Fe is the phase with which
Fe$_{17}$Y$_2$ coexists from melting down to about T=900C, it may be
difficult to observe decomposition of Fe$_{17}$Y$_2$ at low
temperatures where it competes instead with $\alpha$-Fe.  This
scenario suggests the possibility that other Fe-based alloy phase
diagrams could be incorrect at low temperatures, which could have
significance for the engineering of magnetic materials~\cite{RFe}.

The {\em ab-initio} calculations of total energies for the Fe-Zr and
Fe-Y alloys were difficult because of the complicated magnetic
properties of the Fe-rich structures. We mention two important
observations.  (1) The Zr and Y atoms have magnetic moments pointing
opposite to the Fe atoms.  Moments are typically in the range of +1.8
to +2.4 $\mu_B$ for Fe and in the range of -0.2 to -0.4 $\mu_B$ for Y
or Zr in spin polarized calculations. Similar values for Fe-Y alloys
were reported in prior calculations~\cite{Coe89} and
experiments~\cite{Moz94}. (2) Magnetism couples strongly with atomic
volume leading to multiple self-consistent solutions of the
DFT. Generally one finds: a nonmagnetic, low volume, high energy
solution; a strongly magnetic, high volume, low energy solution;
occasional additional solutions of intermediate magnetism, volume and
energy. Presumably this is related to the strong magnetovolume effects
that actually occur in Fe-rich compounds~\cite{RFe,Coe89}.

Comparing our calculated enthalpies of formation with published
experimental data~\cite{LB}, we again find that our calculated values
lie well below the published data, in the case of Fe-rich alloys,
probably as a result of the high temperatures at which the experiments
were carried out. At lower Fe-content our data is fairly consistent with
the experimental data.

\subsubsection{\label{sec:other-bin}Other binaries: B-Zr, B-Y and Zr-Y}

Next we turn to the B-Zr and B-Y binaries. The established B-Zr phase
diagram exhibits three compounds. The well known stable phase
B$_2$Zr.hP3 is very strongly bound.  The other two phases exist only
at high temperatures: B$_{12}$Zr.cF52 melts congruently, while BZr.cF8
exists only over an intermediate temperature range below all liquidus
temperatures~\cite{Glaser,Champion}.  Our calculation
(Fig.~\ref{fig:b-zr}) supports stability of B$_2$Zr and high
temperature stability of B$_{12}$Zr, but strongly contradicts the
existence of BZr.cF8.  Indeed, at this composition the best structure
we find is BCr.oC8, but that too is highly unstable.  We explored
nearby compositions, and even added traces of C (CZr.cF8 {\em is} a
stable compound in the C-Zr binary system), but we cannot find any
structure within a reasonable distance of the convex hull.  Probably
BZr.cF8 is a metastable structure formed during rapid
quench~\cite{Glaser,Champion}.

Comparing our calculated enthalpies of formation with published
data~\cite{LB} we find excellent agreement.  For B$_{12}$Y we
calculate -219 meV/atom compared with the experimental value -213
meV/atom.  The experimental data is taken at a fairly low T=298K.  For
B$_2$Zr we calculate $\Delta H_{for}$=-999 meV/atom compared with the
experimental value -1074 meV/atom.

The enthalpy of B$_2$Zr reflects strong covalent B-B bonding.
Densities of states among transition-metal diborides\cite{AlB2}
exhibit a strong pseudogap associated with Boron p-states. As one
moves across the transition-metal series, the Fermi level falls in the
pseudogap for group IVa elements (Ti/Zr,Hf), leading to strong peaks
in cohesive energy.  This is also the likely cause of the wide
composition range of B$_2$Nb -- since Nb lies just to the right of Zr
and the Fermi energy of $B_2$Nb lies just to the right of the
pseudogap, Nb vacancies can move the Fermi energy closer to the gap,
resulting in a low vacancy formation energy.

In contrast to B-Zr, B-Y contains several B-rich phases, including
B$_2$Y.hP3 and B$_{12}$Y.cF52 as in B-Zr, and additionally B$_4$Y.tP20
and B$_6$Y.cP7.  The last one is of uncertain
composition~\cite{BinaryPD}, with the experimental diagram showing a
composition range at low temperatures, contradicting the notion that
alloys should reach definite compositions as T $\rightarrow 0$K.  We
investigated Yttrium vacancies within a $2 \times 2 \times 2$
supercell of B$_6$Y.cP7 and found that removal of a single Y out of 8
was favorable and lowered the value of $\Delta E$ to +15 meV/atom.
Finally, there is a phase B$_{66}$Y.cF1880 of whose gigantic unit cell size
prevents us from calculating cohesive energy.  Our calculated convex hull
(Fig.~\ref{fig:b-y}) agrees perfectly with the experimental data
except in the case of B$_6$Y which we find is unstable at low
temperature.  Owing to the lack of a definite low temperature
stoichiometry, B$_6$Y most likely {\em is} unstable at low
temperatures~\cite{improb,improb2}.  No experimental data is available
for enthalpies of formation.

Finally, consider the Y-Zr binary (Fig.~\ref{fig:y-zr}). The
established phase diagram contains no compounds. We investigated
several possibilities appropriate for their atomic size ratio and
found none stable.

\subsection{\label{sec:ternary}Ternaries}

\subsubsection{\label{sec:BFeZr+BFeY}B-Fe-Zr and B-Fe-Y}

\begin{figure*}
\includegraphics*[width=5in]{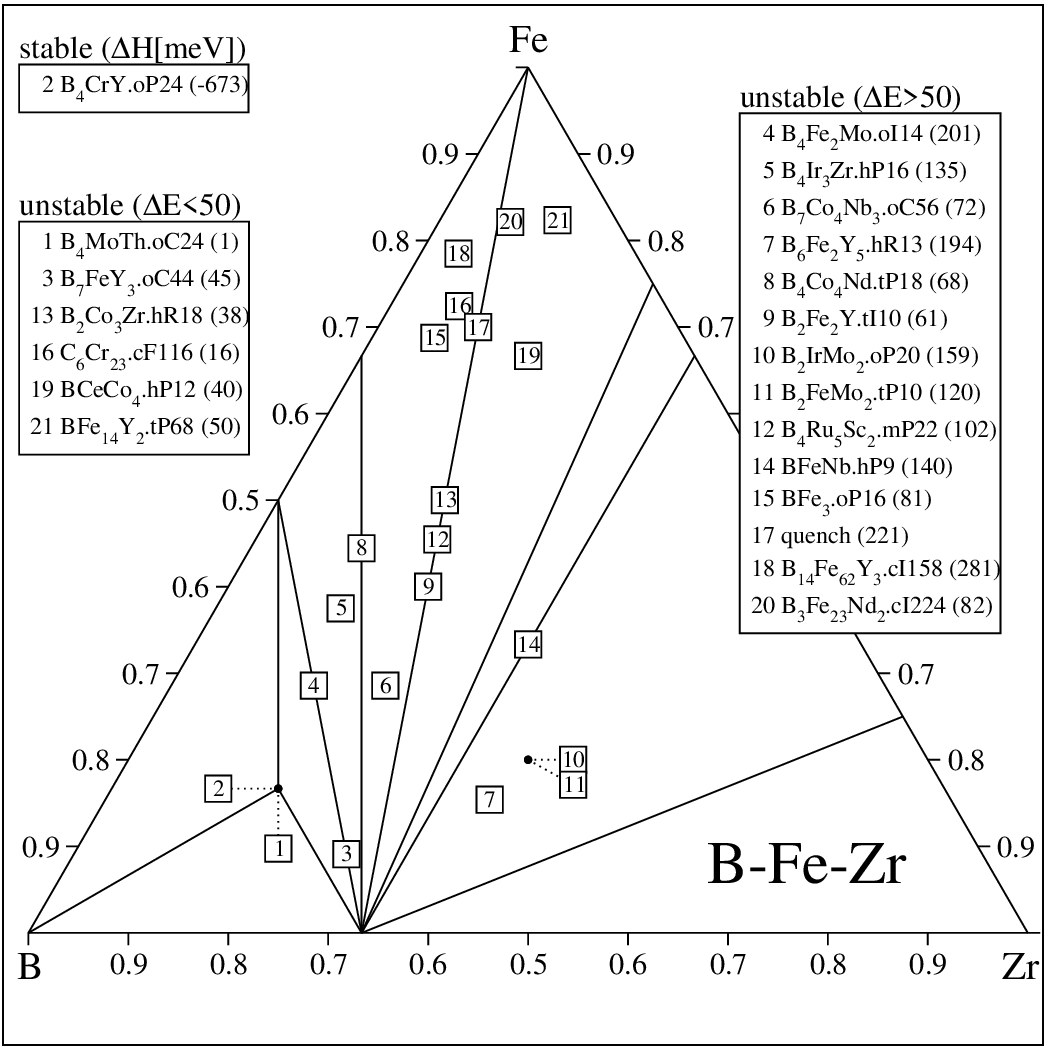}
\caption{\label{fig:b-fe-zr} Convex hull and metastable phases of the
B-Fe-Zr ternary system.  For binary structure types see
Figs.~\ref{fig:b-fe},\ref{fig:b-zr},\ref{fig:fe-zr}.  Plotting symbols
as in Fig.~\ref{fig:b-fe}. Energy units are meV/atom.}
\end{figure*}
\begin{figure*}
\includegraphics*[width=5in]{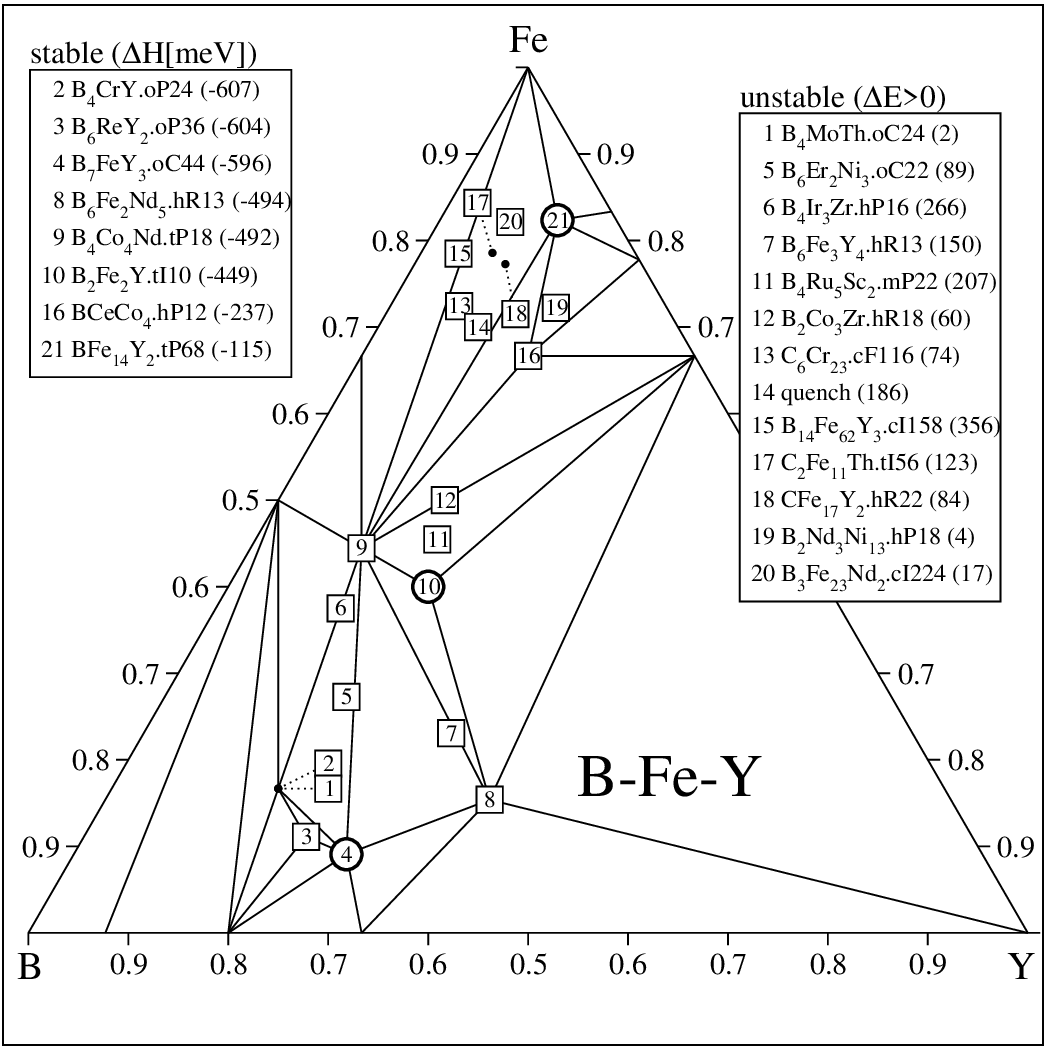}
\caption{\label{fig:b-fe-y} Convex hull and metastable phases of the
B-Fe-Y ternary system.  For binary structure types see
Figs.~\ref{fig:b-fe},\ref{fig:b-y},\ref{fig:fe-y}.  Plotting symbols
as in Fig.~\ref{fig:b-fe}. Energy units are meV/atom.}
\end{figure*}

The ternary alloy system B-Fe-Zr exhibits no known ternary compounds.
Our calculations (Fig.~\ref{fig:b-fe-zr}) generally support this, but
we do identify one stable compound, with structure, B$_4$CrY.oP24.
We convert the binary structure C$_6$Cr$_{23}$.cF116 to a
ternary by substituting the large Zr atoms on sites that have the
largest Voronoi volume.  These turn out to sites of Wyckoff type $8c$,
resulting in the intrinsic ternary structure type
B$_6$Co$_{21}$Zr$_2$.cF116.  Interest in this structure is motivated
by its proximity to the glass-forming composition, and the
metastability of this structure is discussed later in
section~\ref{sec:Discuss}.  The quenched structure is, as before, a
100 atom model metallic glass reached by molecular dynamics and
quenching.

The reason that B-Fe-Zr exhibits so few (i.e. just one) stable ternary
compounds is that the enthalpy of formation of B$_2$Zr is very large,
as discussed above in section~\ref{sec:other-bin}.  Covalent
bonding of B$_2$Zr is so strong that even Fe-rich systems find it
advantageous to phase separate into a mixture of B$_2$Zr plus Fe
alloyed with which ever of Zr or B remains in excess.

In contrast, the B-Fe-Y ternary alloy system (Fig.~\ref{fig:b-fe-y})
exhibits many stable ternary compounds, because the bonding of B$_2$Y
is less strong than B$_2$Zr.  Those compounds with known structures
are the Fe-rich compound BFe$_{14}$Y$_2$.tP68, and in the Fe-poor
region B$_2$Fe$_2$Y.tI10, B$_6$Fe$_3$Y$_4$.hR13, B$_7$FeY$_3$.oC44 and
a metastable structure B$_{14}$Fe$_{62}$Y$_3$.cI158. All previously
known stable structures touch the convex hull. Additionally there are
stable compounds of unknown or partially known structure at
B$_3$FeY$_2$, B$_4$Fe$_4$Y and B$_4$FeY.  Our calculations reveal
these compounds to take the structure types, respectively,
B$_6$Fe$_2$Nd$_5$.hR13, B$_4$Co$_4$Nd.tP18 and B$_4$CrY.oP24.

Surprisingly, we find a previously unknown compound of structure type
BCeCo$_4$.hP12 on the convex hull in the Fe-rich region. Nearby we
find structure type B$_2$Nd$_3$Ni$_{13}$.hP18 just slightly higher in
energy.  It would be of interest to explore these compositions
experimentally in more detail.

At $\sim$80\% B content, we find three stable crystals in the B-Fe-Y
ternary: B$_4$CrY.oP24, B$_6$ReY$_2$.oP36 and B$_7$ReY$_3$.oC44.  One
of these, B$_4$CrY.oP24, is also stable in the B-Fe-Zr ternary.  All
three of these structure types can be considered as approximants to
decagonal quasicrystals.  However, we have not identified systematic
extensions toward truly quasiperiodic structures, and we are not
prepared to predict the occurrence of decagonal quasicrystals in these
compounds.  At present no B-based quasicrystals are known.  Further
discussion can be found in Ref.~\cite{DecApp}.

One structure reported in the B-Fe-Y system,
B$_{14}$Fe$_{62}$Y$_3$.cI158, has a very high energy ($\Delta E=330$
meV/atom) and large initial forces (as high as 1.7 eV/\AA). Even after
large atomic displacements during relaxation the energy remains very
high.  We believe the experimentally reported structure is incorrect.

We are impressed by the faithfulness with which our calculations
reproduce systematic differences in the phase diagrams of Y- and
Zr-containing alloys. Despite their adjacency in the periodic table,
and the consequent similarities in atomic size, electronegativity and
preferred structure types, those details on which the accepted phase
diagrams {\em do} differ are almost always correctly reproduced.

\begin{figure*}
\includegraphics*[width=5in]{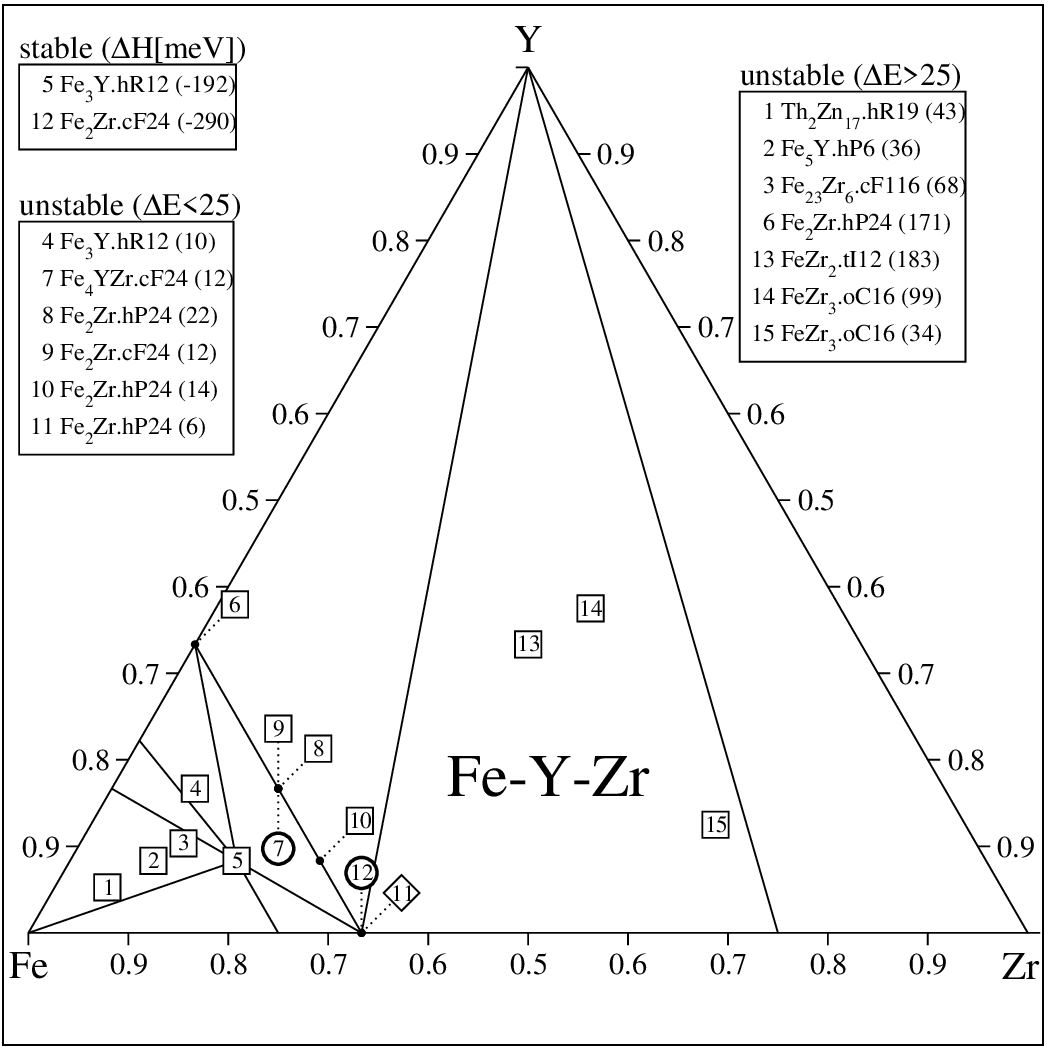}
\caption{\label{fig:fe-y-zr} Convex hull and metastable phases of the
Fe-Y-Zr ternary system.  For binary structure types see
Figs.~\ref{fig:fe-zr},\ref{fig:fe-y},\ref{fig:y-zr}.  Plotting symbols
as in Fig.~\ref{fig:b-fe}. Energy units are meV/atom.}
\end{figure*}

\begin{figure*}
\includegraphics*[width=5in]{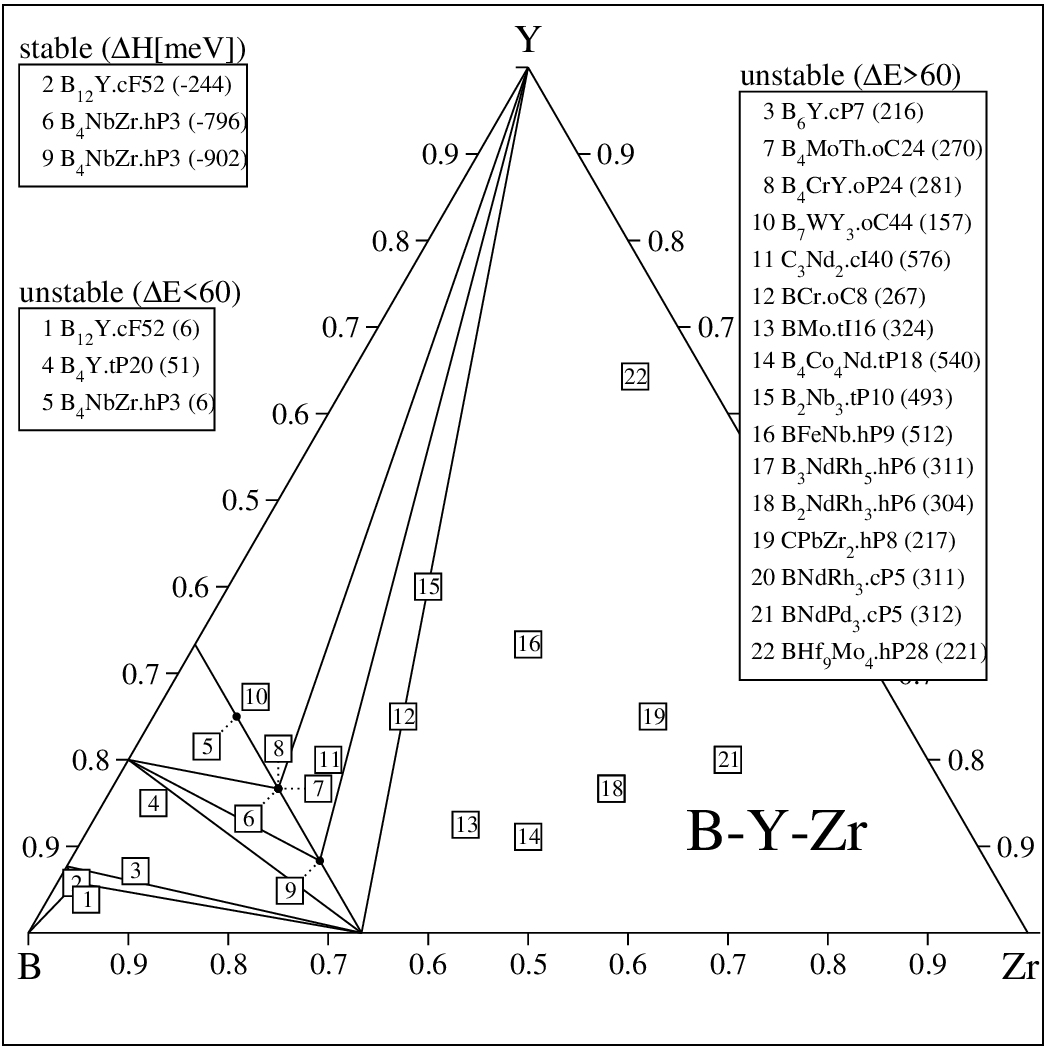}
\caption{\label{fig:b-y-zr} Convex hull and metastable phases of the
B-Y-Zr ternary system.  For binary structure types see
Figs.~\ref{fig:b-zr},\ref{fig:b-y},\ref{fig:y-zr}.  Plotting symbols
as in Fig.~\ref{fig:b-fe}. Energy units are meV/atom.}
\end{figure*}

\subsubsection{\label{sec:other-tern}B-Y-Zr and Fe-Y-Zr}

The B-Y-Zr ternary diagram has not been experimentally determined. We
have explored it using the methods described above. The only stable
ternary compounds we find (Fig.~\ref{fig:b-y-zr}) are extensions of
certain binaries into the ternary. Notably B$_2$(Zr,Y).hP3 exhibits
complete miscibility of Zr and Y in this pseudobinary
structure. Additionally, B$_{12}$Y.cF52 extends part way into the
ternary.

The Fe-Y-Zr ternary diagram has not been experimentally determined
either.  Our calculations (Fig.~\ref{fig:fe-y-zr}) suggest that
Fe$_2$(Y,Zr).cF24 and Fe$_3$(Y,Zr).hR12 both extend across the full
ternary diagram, but no other binaries appear to extend far into the
ternary.

\subsection{\label{sec:quaternary}B-Fe-Y-Zr Quaternary}

The B-Fe-Y-Zr quaternary has not been experimentally determined. No
quaternary structures are reported in the standard references. We have
calculated enthalpies of formation for 15 different compounds (11
structure types, some with alternate chemical occupancies) and find no
stable quandaries.  Our lowest energy structures are listed in
Table~\ref{tab:quaternary}. The nearest we come to stability is for
the structure type B$_4$CrY.oP24, for which $\Delta H_{for}$ is around
3-4 meV/atom for all substitutions of Y and Zr. Thus it is likely that
the entire Y/Zr substitution yields equilibrium structures at high
temperatures. We find 4 meV/atom for substitution of Zr for one of the
two Y in BFe$_{14}$Y$_2$.tP68, suggesting significant Zr solubility at
high temperatures.

Several factors contribute to the lack of stable B-Fe-Y-Zr
quandaries: incompatibility of Y and Zr atoms (see
Fig.~\ref{fig:y-zr}) destabilizes quaternaries that are rich in Y or
Zr; the difficulty of accommodating the slightly differing atomic sizes
into the same crystal lattice site classes destabilizes quaternaries in
which Y and Zr are minority species; the very strong bonding of B with
Zr destabilizes quaternaries that are B-Zr-rich.

\begin{table}
\caption{\label{tab:quaternary}Quaternary data}
\begin{tabular}{|r|r|r|r||r|r||l|l|}
\hline
 B & Fe &  Y & Zr & $\Delta E$ & $\Delta H_{for}$ & Structure & Comments\\
\hline
67 & 17 &  4 & 12 &   3.1 & -653 &B$_4$CrY.oP24 & 3Zr \\
67 & 17 &  8 &  8 &   4.4 & -636 &B$_4$CrY.oP24 & 3Zr \\
67 & 17 & 12 &  4 &   3.8 & -620 &B$_4$CrY.oP24 & 1Zr \\
 7 & 82 & 10 &  1 &   5.2 & -117 &BFe$_{14}$Nd$_2$.tP68 & 1Zr \\
 7 & 82 &  6 &  6 &  21.9 & -120 &BFe$_{14}$Nd$_2$.tP68 & 4Zr \\
64 &  9 & 23 &  4 &  21.1 & -623 &B$_7$FeY$_3$.oC44 & Zr on 4c \\
64 &  9 & 23 &  4 &  28.4 & -620 &B$_7$FeY$_3$.oC44 & Zr on 8f \\
17 & 67 &  8 &  8 &  23.2 & -264 &BCeCo$_4$.hP12 & \\
21 & 72 &  3 &  3 &  45.9 & -210 &C$_6$Cr$_{23}$.cF116 & Y/Zr on 8c\\
20 & 70 &  5 &  5 & 211.9 &  -61 &quench & \\
\hline
\end{tabular}
\end{table}

\section{\label{sec:Discuss}Discussion}

\subsection{\label{sec:Benv}Boron atom environments}

\subsubsection{Trigonal prisms in Fe-rich B-Fe stable and metastable systems}

Trigonal prisms~\cite{Gaskell} place six large (Fe) atoms at their six
vertices, their rectangular (nearly square) faces are capped by an
additional three large atoms, and they are centered by a small (B)
atom. They are well known structural motifs in compounds with
significant contrast in atomic size, in the large-atom-rich
composition range.  Too large a size contrast is unfavorable,
as they do not occur in B-Y or B-Zr binaries.  A stringent definition
of the trigonal prism uses the (radical-planes) Voronoi construction.
The Voronoi polyhedron of the central atom should have no triangular
face, 3 rectangular and 6 pentagonal faces. This polyhedron is denoted
(0,3,6) in the ($n_3$,$n_4$,$n_5$...) notation of Watson and
Bennett~\cite{WB}.

One stable B-Fe compound, BFe.oP8 which we find marginally more stable
than the BCr.oC8 prototype, contains trigonal prisms.  In both
structures (and also in closely related CaCu.mP20 and CaCu.oP40
structure types) Fe prisms share 2 out of 3 rectangular faces with
neighboring prisms, while the third rectangular face is capped by an
Fe atom.  The prisms form columns along the shortest periodic
direction.  The structure has respectable packing fraction (greater
than 0.73) when Fe/B atoms are replaced by hard spheres with radius
ratio 1.55, optimizing the packing fraction.

In the Fe-rich portion of the B-Fe system (see
section~\ref{sec:Fe-B}), we find seven Fe-rich structures that are
unstable by less than 50 meV/atom relative to the convex hull.  These
include the known metastable phases BFe$_3$.oP16, BFe$_3$.tI32 and
B$_6$Fe$_{23}$.cF116, and other structures BRe$_3$.oC16,
C$_2$Fe$_5$.mC28, C$_3$Fe$_7$.oP40 and C$_3$Fe$_7$.hP20. Of these, the
BRe$_3$.oC16 and CoSc$_3$.oP32 structures are sufficiently low in
energy and differ sufficiently in composition from the nearest stable
crystalline phases that we expect they could also occur as metastable
phases. At higher B-content we find B$_3$Ni$_4$.oP28 at low energy and
possibly metastable.

With the exception of the cF116 structure, all B atom environments in
the above-mentioned structures are proper trigonal prisms.  In the
oP16 and oC16 structures all Fe atoms are structurally similar, each
with 3 B neighbors. The oC16 structure is characterized by a unique
stacking mode of the trigonal prisms, forming unterminated columns along
the shortest-period $a$-axis and sharing triangular faces.  Each Fe
has 2 B neighbors, and each Fe is simultaneously the vertex of one
prism and a capping atom of another prism, shifted by $a$/2.  The
B$_3$Ni$_4$.oP28 structure combines the same building blocks found in
BRe$_3$.oC16 and BFe.oP8 - columns of trigonal prisms stacked into
columns either sharing triangular faces (oC16) or rectangular faces
(oP8).

The stable phase BFe$_2$, like a number of other B-TM$_2$ binary
systems, crystallizes into the Al$_2$Cu.tI12 structure type.  Viewed
parallel to the shortest ($c$) axis, the structure is built by two
flat layers of Fe atoms each forming a nearly regular square-triangle
tiling pattern~\cite{DecApp}.  Boron atoms occupy interstitial
octahedral sites in the network.  This topology is not optimal for
packing atoms with very different sizes, so the atomic sizes must not
play an important energetic role for this system.  Interestingly, this
is the only structure among the stable and metastable B-Fe compounds
(with the exception of cF116) in which B atoms do not have
trigonal-prismatic environment.  Each B atom has two other B atoms
only 2.1A distant, forming 1D chains along the $c$-direction.

B atoms in the cF116 structure have the (0,5,4) Voronoi polyhedron; if
we eliminate its smallest face (area 0.35\AA$^2$), it converts to an
(0,8,0) environment, in which each B atom has 8 Fe neighbors.  This B
environment is similar to that of BFe$_2$ except there are no B-B near
neighbors.  The cF116 structure is also exceptional in an uneven
distribution of the B atoms in Fe matrix: while some Fe atoms (sites
4a and 8c) have no B neighbors, site 32f has two B neighbors and 48h
has three.

\subsubsection{Quenched samples of Fe$_{80}$B$_{20}$}

In the quenched samples, relaxed to the local minimum in cohesive
energy at 0K, the most common B environments are: (i) trigonal prisms
(0,3,6) with 9 Fe atom environments; (ii) the (0,5,4), with 8 Fe and 1
B atom environments. These occur in roughly equal proportion.  The
trigonal prisms are consistent with the main structures of the Fe-rich
metastable phases, while the (0,5,4) environments are characteristic of
the stable BFe$_2$.tI12 structure in which B-B neighbors occur. In
fact we occasionally find a B atom with two B neighbors, resulting in
local environments very close to BFe$_2$.

Diffraction data~\cite{Nold} finds no B-B neighbors, but they occur
robustly in our simulations and perhaps can serve as nucleation sites
for crystalline BFe$_2$. There is some controversy in the literature
about the certainty with which B-B neighbors can be ruled
out~\cite{Cowlam}.

\subsubsection{Boron environments in Fe-rich ternaries}

For B-Fe-Y, in the stable compounds BCeCo$_4$.hP12 and
BFe$_{14}$Y$_2$.tP68, as well as B$_2$Nd$_3$Ni$_{13}$.hP18 structure
(which in our calculation is unstable by just 4meV/atom), the B
environment is a trigonal prism with Fe at vertices and Y capping the
rectangular faces.  Interestingly, in the metastable
B$_3$Fe$_{23}$Nd$_2$.cI224 (this lies just 12 meV/atom above the
tieplane) the trigonal prisms come in pairs, sharing rectangular faces
and creating one B-B bond per pair.  In contrast, in the cF116
structure, which is nearly stable in the B-Fe-Zr system, B atoms are
surrounded by Fe atoms only.

\subsection{\label{sec:Fe17Y2}Structure of Fe$_{17}$Y$_2$}

The structure of the compound Fe$_{17}$Y$_2$ is not precisely known.
Multiple structural variants have been observed, and the best
structure refinements contain many partially occupied sites.  The
structures have close structural relationship~\cite{RFe} to
Fe$_5$Y.hP6 (CaCu$_5$ prototype), in which Y atoms center hexagonal
columns of Fe atoms. The columns of Y atoms ($c=4.1$\AA) form a
triangular lattice with edge length $a=4.9$\AA.  Starting from this
structure, the Fe$_{17}$Y$_2$ family may be derived by (i) taking the
superstructure defined by the vectors
$(1,-1,0)\times(1,2,0)\times(0,0,2)$; (ii) applying the substitutional
rule Y$\rightarrow$2Fe.  Neighboring Y atoms (separated by either $a$
or $c$ distances) should never be substituted simultaneously.  This rule
enforces planar hexagonal lattices of Y atoms, with apparent stacking
degrees of freedom~\cite{Moz94}.  The Th$_2$Zn$_{17}$.hR19 prototype
(also known as $\alpha$) takes the $ABC$ stacking sequence (we denote
4\AA~ bilayers by capital letters) with $c_{\alpha}\sim 12$\AA. The
Ni$_{17}$Th$_2$.hP38 prototype (also known as $\beta$) takes the $AB$
sequence with $c_{\beta}\sim 8$\AA.  The two reported refinements of
Fe$_{17}$Y$_2$ (hP44~\cite{hP44} and hP80~\cite{hP80}) are apparently
disordered versions of $\beta$-Ni$_{17}$Th$_2$.

Our calculation confirms small energy differences between the stacking
variants: we find $AB\equiv AC$ is 0.9 meV/atom higher in energy than
$ABC$, which in turn is 0.9 meV/atom higher in energy than the the
$ABAC$ sequence with $\sim$16\AA~ stacking period. The $ABAC$
sequence, which is the best we have found, leads to fractional
occupancy of some Wyckoff positions in qualitative agreement with the
hP44 refinement~\cite{hP44}.

Experimentally reported fractional occupancies, along with occupancies
of the stacking sequences we studied, are reported in
Table~\ref{tab:fe17y2}.  The Y$\rightarrow$2Fe substitution rule
together with our assumption of a disordered ``$A*A*$'' stacking
sequence, constrains site occupancies $p$ so that $p$(Y1)+$p$(Y2)=1,
and $p$(Fe1)+$p$(Fe2)=1. Thus the hP44 refinement implies some Y
vacancies, while the hP80 refinement places extra Fe atoms at Fe1+Fe2. We
considered these possibilities, but found both of them energetically
unfavorable.  Therefore we believe that Fe$_{17}$Y$_2$ is the correct
stoichiometry, and the mismatches in occupancy factors are artifacts
of the refinement, arising from stacking disorder.

The absence of well-ordered crystalline samples further supports our
proposal that the Fe$_{17}$Y$_2$ phase could be unstable at low
temperatures.

\begin{table}
\caption{\label{tab:fe17y2} Fractionally occupied Wyckoff sites in
Fe$_{17}$Y$_2$ structures.  First column labels sites as in the hP44
refinement~\cite{hP44}, second column ($\mu$) gives the number of
equivalent atoms per unit cell. Stackings $AB$ and $AC$ are different
crystallographic settings, but otherwise completely equivalent.
Fe$_\alpha$ sites appear when we register the hR19 structure of
Th$_2$Zn$_{17}$ ($ABC$ stacking) with the hP38, hP44 and hP80
structures.  This atom is not present in either hP44 or hP80
refinements. The final row reports $\Delta E$ in units of meV/atom.}
\begin{tabular}{cc|rrrrrr}
site&$\mu$&  $AB$&   $AC$&  $ABAC$& $ABC$&  hP44&  hP80\\
\hline
Y1  &   2&   1&  1/2&   3/4&  2/3&  0.41&  0.71\\
Y2  &   2&   0&  1/2&   1/4&  2/3&  0.35&  0.12\\
Y3  &   2&   1&    1&     1&  2/3&  1.0 &  1.0 \\
Fe1 &   4&   0&  1/2&   1/4&  1/3&  0.28&  0.29\\
Fe2 &   4&   1&  1/2&   3/4&  1/3&  0.71&  0.86\\
Fe$_\alpha$&4&0&   0&     0&  1/3&     0&  0\\
\hline
$\Delta E$ & & 21.4 & 21.4 & 20.7 & 21.0 & & \\
\end{tabular}
\end{table}

\subsection{\label{sec:glass}Glass formation}

By inspection of our cohesive energy data we can identify the main
crystal phases that are likely to compete with formation of the
amorphous solid. In the vicinity of Fe$_{80}$B$_{20}$, the structure
C$_6$Cr$_{23}$.cF116 can crystallize with almost no composition
shift. However, it may be difficult to nucleate and grow such a
complex crystal type during a quench, so a simpler nearby structure
such as Fe$_3$B.oP16 may be favored. In the limit of slow cooling,
phase separation into pure Fe and Fe$_2$B will occur. In fact, all
these structures are reported in annealed samples of Fe-B glasses.

Alloying with Zr can be advantageous because the large and strongly
interacting Zr atoms diffuse slowly~\cite{MRS02}. Due to its large
size, Zr strongly destabilizes the Fe$_3$B.oP16 structure.  However,
we see in table~\ref{tab:enth} that Zr actually tends to stabilize
slightly the C$_6$Cr$_{23}$ structure, and also risks formation of the
BFe$_{14}$Nd$_2$.tP68 structure. Stabilization of the C$_6$Cr$_{23}$
structure may be counteracted with replacement of Y for Zr.

Another danger of alloying with Zr is formation of the highly stable
binary B$_2$Zr.hP3. Indeed, in the B-Fe-Zr ternary~\cite{TernaryPD}
even Fe-rich liquids coexist with solid B$_2$Zr.  Choosing
compositions with B content below the eutectic (17\% B) can avoid
B$_2$Zr formation.  Also, alloying with the less strongly interacting
(but still very large) element Y can counteract this.  However, it is
not advantageous to alloy only with Y because 1) Y atoms diffuse more
quickly than Zr as a result of their weaker binding; 2) There are
several Fe-rich Fe-Y binary structures (e.g. Fe$_5$Y, Fe$_{17}$Y$_2$)
whose formation should be avoided.  A reasonable composition that
balances these difficulties is B$_{15}$Fe$_{75}$Y$_3$Zr$_7$.

\begin{table}
\caption{\label{tab:enth}Enthalpies of crystal phases competing
with glass formation.}
\begin{tabular}{|l||r|r|r|r|}
\hline
  & Fe$_3$B.oP16 & C$_6$Cr$_{23}$.cF116 & BFe$_{14}$Nd$_2$.tP68 & quench\\
\hline
B-Fe    &  -211 (17) &   -168 (18)  &  +215 (268)   & -29 (151) \\
B-Fe-Y  &  -27 (253) &   -163 (74)  &  -115 (0)     & -59 (186) \\
B-Fe-Zr &  -218 (81) &   -253 (16)  &  -116 (50)    & -79 (221) \\
\hline
\end{tabular}
\end{table}

Recent experimental studies~\cite{Lu03,Poon03} of Y addition to
Fe-Zr-B-based glasses confirm its beneficial effects, while attributing
them to different causes~\cite{Lu03}.

\subsection{\label{sec:Approx}Canonical-cell models of amorphous approximants}

Canonical cell tilings~\cite{CLH} (CCT) form networks of icosahedral
cluster centers for models of icosahedral quasicrystals.  They were
motivated by the cubic ``1/1'' approximants of the quasicrystals, in
which icosahedral clusters are located at the vertices of BCC lattice,
and connected by 2-fold (``$b$'') and 3-fold (``$c$'') inter-cluster
linkages. Linkages of $b$-type are longer than $c$-type linkages by
the factor 2/$\sqrt{3}$=1.15.  The $bc$-network with global
icosahedral symmetry and maximal density of clusters is a tiling of
four canonical cells, $A$, $B$, $C$ and $D$. An $A$-cell is the
two-fold symmetric BCC-tetrahedron, $B$-cell is a skewed rectangular
pyramid, $C$ cell is a three-fold symmetric tetrahedron and $D$-cell
is a trigonal prism.  The trigonal prism has equilateral triangular
bases of $b$-type linkages and rectangular faces of $b$- and $c$-type
linkages.

Consider a model amorphous structure in which the icosahedral clusters
are replaced by ``large'' atoms, and the largest of the cells, the
trigonal prism $D$, is decorated by a small atom in the center.  Such
a model appears to be entirely plausible for B-Fe, since: (i) ideal B
atom environments in the $D$-cell produce very nearly the same ratio
of B-Fe/Fe-Fe nearest neighbor distances as the BFe$_3$ compounds;
(ii) provided the density of $D$-cells is fixed by B atom content, the
canonical cells models should yield optimal Fe-Fe connectivity via the
$b$ and $c$ linkages, with $b$/$c$ length ratio fixed at the BCC
structure value 2/$\sqrt{3}$. In the limit of large unit cell size
approximating an icosahedral quasicrystal, the B content of this model
ranges from $x_B~\sim 0.11$ to 0.22, covering the best glass-forming
composition range.

As a convenient and simple example of the CCT model, we consider
smallest cubic approximant model containing all kinds of cells, so
called ``3/2'' tiling with 32 CCT nodes per cubic cell, and $Pa\bar 3$
space group.  It contains 8 $D$-cells centered by B atoms (also 72
$A$-cells, and 32 each of $B$- and $C$-cells), and has composition Fe$_4$B.
This ideal model remains practically undistorted upon relaxation, and
relaxes to an energy $\sim$210 meV/atom above the tie-line.  B-Fe and
Fe-Fe bond lengths in the model are similar to those found in the
metastable crystalline compounds of similar composition.  Performing
Voronoi analysis we find that the Fe atom polyhedron volume is similar
to that of the crystals, but the B atom polyhedron volume is larger.
In the CCT model the B atom polyhedron volume is 5.03 \AA$^3$, similar
to the volume of 4.9\AA$^3$ in relaxed amorphous structures but larger
than the 4.6-4.7\AA$^3$ occurring in the metastable crystalline phases.
Because the energies and geometry of the CCT models are close to the
amorphous structure, we call the CCT models ``amorphous
approximants''.

\section{Conclusions}

In conclusion, we present a method for the calculation of low
temperature (T=0K) alloy phase diagrams and apply it to the study of
the B-Fe-Y-Zr quaternary system. A key distinguishing feature of our
approach is the establishment of two databases: one from which we draw
promising structures observed in similar chemical systems; the other in
which we record our calculated cohesive energies which can be quickly
converted into enthalpies of formation. The first database allows us
to propose and evaluate candidate structures even in alloy systems
that have not been previously studied. The second allows us to quickly
add a new chemical element, and re-use, for example, all our B-Fe-Zr
data in the study of the B-Fe-Y-Zr quaternary.

As a result we have examined two new ternary systems (B-Y-Zr and
Fe-Y-Zr) and the B-Fe-Y-Zr quaternary. We found certain binary phases
extending into the new ternary systems, but no ternary phases
extending into the quaternary. So far we have not discovered any stable
quaternary structure.  Even in previously studied binary and ternary
systems we find some new results including proposed structures for
previously unsolved compounds.

The broad agreement between our calculations and experimentally
reported phase diagrams demands that special attention be paid where
disagreements exist. These disagreements fall into certain categories:
Uncertain reported structure found to have high energy (e.g. BZr.cF8);
Well established experimental structure found to have high energy
(e.g. Fe$_{17}$Y$_2$); Structure calculated to be stable not present
in published diagram (e.g. B$_4$CoZr in B-Fe-Zr and Co$_7$Y$_2$ in
Fe-Y). These disagreements warrant further study, both theoretical and
experimental.

On the subject of glass formation, the main motivation for this study,
we identify important crystalline competitors to glass-formation and
illustrate how they can be destabilized by the addition of
appropriately chosen large atoms.

\begin{acknowledgments}
We wish to acknowledge useful discussions with Libo Xie, Yang Wang, Don
Nicholson, Joe Poon and Gary Shiflet. This work was supported by
DARPA/ONR Grant N00014-01-1-0961.
\end{acknowledgments}

\bibliography{fezryb4}

\end{document}